# Molecular Doping of Multilayer MoS$_2$ Field-Effect Transistors: Reduction in Sheet and Contact Resistances

Yuchen Du, Han Liu, Adam T. Neal, Mengwei Si, and Peide D. Ye, *Fellow, IEEE*

*Abstract*—For the first time, polyethyleneimine (PEI) doping on multilayer MoS$_2$ field-effect transistors is investigated. A 2.6 times reduction in sheet resistance and 1.2 times reduction in contact resistance have been achieved. The enhanced electrical characteristics are also reflected in a 70% improvement in ON-current and 50% improvement in extrinsic field-effect mobility. The threshold voltage confirms a negative shift upon the molecular doping. All studies demonstrate the feasibility of PEI molecular doping in MoS$_2$ transistors and its potential applications in layer-structured semiconducting 2-D crystals.

## I. Introduction

TWO-DIMENSIONAL layered-structure materials have attracted much interest in recent years for their unique physical properties in thermal, optical, and electronic devices. One of the pioneering materials, graphene, is noteworthy for its large mobility up to $10^6$ cm$^2$/V·s [1]–[3]. The absence of a bandgap in graphene, however, prohibits its use in logic applications [4]. Alternatively, semiconducting transition metal dichalcogenides (TMDs) material has been recognized as another type of 2-D material for possible solid-state device applications [5]–[11]. MoS$_2$, a typical layer-structured TMD family material has enjoyed many advantages in device applications because it usually has large bandgaps ($>$1 eV) [12] and reasonable carrier mobility [8], [13], [14]. Besides, its 2-D nature shows the immunity to short-channel effects at reduced device dimensions [15], [16], as well as mechanical flexibility. Chemical sensors, photonic detectors, memories, and integrated circuits have been widely demonstrated on single or few layers MoS$_2$ [17]–[21]. In addition, all those devices are based on individual MOSFETs, giving more demands on single transistor performance. To improve the device performance, recent research on MoS$_2$ field-effect transistors (FETs) has been mostly focusing on the following issues: 1) scaling down the device dimensions [22]; 2) top-gating devices with high-$k$ materials [23], [24]; 3) reducing contact resistance [25]; and 4) studying MoS$_2$ FETs transport behavior on different substrates [14]. Engineering electronic performance via doping is, however, still in its infancy for layered crystals, due to its ultrathin body nature. MoS$_2$ may not be doped as silicon by traditional methods such as ion implantation; however, the ultrathin body nature allows the exploration of novel approaches, such as chemical and molecular doping. Previous research of chemical doping method demonstrates the surface charge transfer between potassium and MoS$_2$ [26]. In this letter, we present a simple approach to incorporate doping in multilayer MoS$_2$ flakes by adopting polyethyleneimine (PEI) molecular as dopants. Our research sheds lights on the reduction in both sheet and contact resistances via molecular doping, fully demonstrates the charge-transfer enabled molecular doping of MoS$_2$.

## II. Experiments

The amine-rich aliphatic polymer, PEI (Sigma Aldrich, $M_n \sim 60\,000$, and $M_w \sim 750\,000$), shown in Fig. 1(a), is a widely used $n$-type surface dopant, for doping low-dimensional nanomaterials devices due to its strong electron-donating ability, ease of application, and large molecular weight [27]–[30]. A small amount of PEI in methanol ($\sim$0.02 wt%) was prepared by magnetic stirring for 48 h and the PEI solution was stored in a dark, air-tight container for future usage. MoS$_2$ flakes with an average thickness of 4–5 nm were mechanically exfoliated from bulk ingot (SPI Supplies) by standard scotch tape technique, and then transferred to a heavily p-doped silicon substrate with a 90-nm SiO$_2$ capping layer. After the flake transfer onto the SiO$_2$/Si substrate, electron-beam lithography was used to pattern the source and drain contacts, where the width of contact bars was 1 $\mu$m. A transmission line method (TLM) structure has been patterned with channel length of 500 nm, 1, 2, and 3 $\mu$m. Metallization was performed by electron-beam evaporation of 20/60 nm (Ti/Au) with a background pressure of $10^{-6}$ Pa. The TLM structure geometry of MoS$_2$ FETs is shown in Fig. 1(b). Electrical measurements were carried out with Keithley 4200 semiconductor parameter analyzer and probe station in ambient atmosphere. To dope the device after initial characterization, the MoS$_2$ FETs device was soaked in PEI solution for 24 h, followed by a 30-s methanol rinse and N$_2$ gun blow dry. The surface roughness is $\sim$1.6 nm on pristine MoS$_2$ flakes and increases to 5.6 nm after PEI soaking, determined by atomic force microscope (AFM).

Manuscript received June 5, 2013; revised July 15, 2013; accepted July 24, 2013. The work was supported in part by the National Science Foundation under Grant CMMI-1120577 and in part by Semiconductor Research Corporation under Tasks 2362 and 2396. The review of this letter was arranged by Editor Z. Chen.

The authors are with the School of Electrical and Computer Engineering and Birck Nanotechnology Center, Purdue University, West Lafayette, IN 47907 USA (e-mail: yep@purdue.edu).

Color versions of one or more of the figures in this letter are available online at http://ieeexplore.ieee.org.

Digital Object Identifier 10.1109/LED.2013.2277311

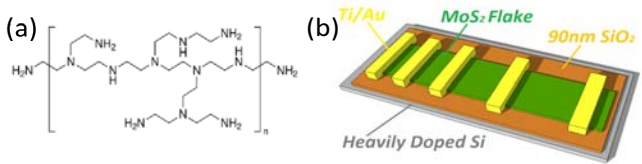

Fig. 1. (a) Molecular structure of PEI. (b) Geometry of MoS$_2$ FETs TLM structure with channel length of 500 nm, 1, 2, and 3 $\mu$m.

## III. RESULTS AND DISCUSSION

Contact resistance and sheet resistance have been extracted from the TLM structure by

$$R_{\text{tot}} = 2R_C + R_s \frac{L}{W} \quad (1)$$

where $R_{\text{tot}}$ is the total measured resistance, $R_c$ is the contact resistance, $R_s$ is the sheet resistance, the channel length $L$ varies from 500 nm to 3 $\mu$m, and the channel width $W$ is 2.5 $\mu$m.

Transfer characteristics of $R_c$ and $R_s$ versus back-gate $V_{\text{bg}}$ before and after PEI doping are shown in Fig. 2(a) and (b). Both sheet resistance and contact resistance show a strong dependence on the back-gate bias due to increasing carrier concentration in the MoS$_2$ flake under positive gate bias. Considering $R_c$, the increased carrier concentration in MoS$_2$ results in a lowering of the effective Schottky barrier height, thus lowering $R_c$. Error bars are determined from the standard errors of the linear fitting under different back-gate bias with changing . Notably, the error bars of both resistances under the lower back-gate bias are much larger than those at high back-gate bias regions, which is due to the larger $R_s$ and $R_c$ values under the low gate bias. The large absolute error bars are also observed in former MoS$_2$ TLM research [22]. To have more accurate comparisons in $R_s$ and $R_c$ over PEI doping method, we choose low absolute error regions at $V_{\text{bg}} = 60$ V. In Fig. 2(a), $R_s$ measured after PEI doping is $7.65 \pm 1.81$ k$\Omega$/□ at 60 V back-gate bias, decreased from $19.99 \pm 3.31$ k$\Omega$/□ before doping. This nearly 2.6 times reduction in $R_s$ occurs from PEI molecular doping treatment due to charge transfer of electrons from the PEI to the MoS$_2$ flake, where PEI acts like an electron donor. $R_c$ varies as a function of back-gate bias is shown in Fig. 2(b), where $R_c$ measured before and after PEI treatment at $V_{\text{bg}} = 60$ V are $5.06 \pm 1.70$ $\Omega$·mm and $4.57 \pm 1.08$ $\Omega$·mm, respectively. Approximately 20% lowering of $R_c$ can be attributed to the reduction of Ti-MoS$_2$ Schottky barrier width for electron injection.

The adjustment in threshold voltage, $V_T$ of various MoS$_2$ transistors by PEI doping method has been observed and quantitatively measured. The $V_T$ is extracted by referring linear extrapolation method with the drain current measured as a function of gate voltage. Before PEI doping, the as transferred MoS$_2$ transistor exhibited an $n$-type transport behavior with threshold voltage, $V_T = -4.51$ V for 3-$\mu$m channel device. Upon application of PEI, the threshold voltage of the doped MoS$_2$ transistor shifted to $-28.59$ V, where the negative threshold voltage shift confirms the strong $n$-type dopant of PEI molecules had been applied on MoS$_2$ FETs. To have further understanding of PEI doping effect on threshold voltage adjustment, before PEI dopant applied, all devices with different channel lengths have been measured and analyzed, and they all demonstrated the same negative shift in $V_T$, indicating the $n$-type doping ability of the PEI, and demonstrating PEI doping could be a possible method to adjust threshold voltage in future 2-D materials CMOS logic circuit. An $I_{\text{ds}}$ versus ($V_{\text{bg}}$–$V_T$) transfer curve of the 3-$\mu$m channel length device is shown in Fig. 2(c) so as to demonstrate the transfer characteristics of a back-gated MoS$_2$ FETs. Before the PEI doping, the ON-current of the device is 10.25 mA/mm at $V_{\text{bg}}$–$V_T = 30$ V. After the PEI doping, MoS$_2$ FETs ON-current increases to 17.61 mA/mm. The 70% enhancement in ON-current is comparable with that of PEI doped graphene FETs [31], which also results from a lowering of the $R_s$ and $R_c$. Extrinsic field-effect mobility before and after PEI doping are calculated to be 20.4 and 32.7 cm$^2$/V·s for 3-$\mu$m channel length device, which are extracted from the $I$–$V$ transfer curves by the relation

$$\mu = \frac{dI_{\text{ds}}}{d(V_{\text{bg}} - V_T - \frac{V_{\text{ds}}}{2})} \times \frac{L}{W \times C_{\text{ox}} \times V_{\text{ds}}} \quad (2)$$

where $dI_{\text{ds}}/d(V_{\text{bg}} - V_T - V_{\text{ds}}/2)$ is the transconductance, $L$ is the channel length, $W$ is the channel width, and $C_{\text{ox}}$ is the back-gate capacitance/unit area. Despite the improvement of ON-current and extrinsic field-effect mobility on MoS$_2$ FETs by doping with PEI solution, a decreased ON/OFF ratio is observed in our experiment. Before PEI doping, the device showed a high ON/OFF ratio of $\sim 10^5$ in ambient atmosphere. After PEI applied, the ON/OFF ratio is, however, reduced to $\sim 10^2$. It could be ascribed to the reduced electrostatic control of the multilayer MoS$_2$ channel by the back gate [32] because the PEI only dopes the top layers by charge transfer. It could also be simply due to the current leakage through the global PEI layer [33], which suggests a localized doping or device isolation after PEI is needed in future research. Fig. 3(d) shows the output characteristics of MoS$_2$ FETs before PEI doping. The improvement of current at same back-gate bias demonstrates that PEI doping successfully improves the performance of MoS$_2$ $n$-FETs.

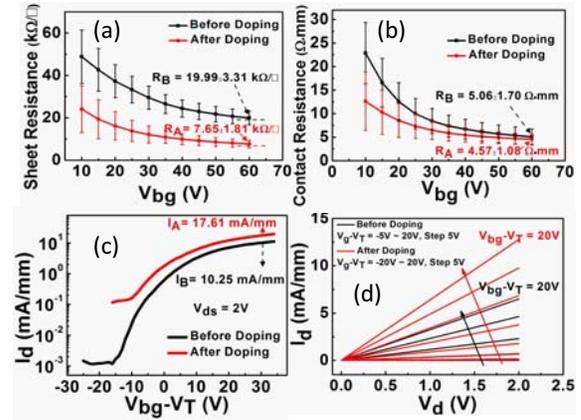

Fig. 2. (a) Comparison of sheet resistance before and after PEI doping varies with different back-gate bias. (b) Comparison of contact resistance before and after PEI doping varies with different back-gate bias. (c) Transfer characteristics of a typical transistor with channel length $L = 3$ $\mu$m before and after PEI doping, with $V_{\text{ds}} = 2$ V. (d) Output characteristics of the same transistor before and after PEI doping. Before doping, $V_{\text{bg}}$–$V_T$ varies from $-5$ to 20 V, with a 5 V step. After doping, $V_{\text{bg}}$–$V_T$ varies from $-20$ to 20 V, with a 5 V step.

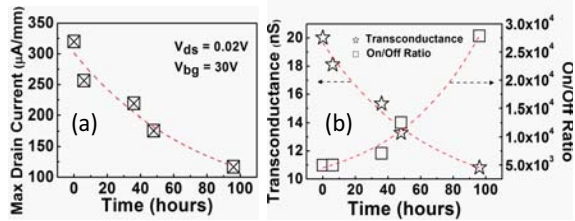

Fig. 3. (a) Maximum drain current. (b) Peak transconductance and ON/OFF ratio as a function of time exposed in ambient air at $V_{ds} = 0.02$ V.

The 3-$\mu$m channel length MoS$_2$ transistor had been stored and measured repeatedly in ambient atmosphere. The long term air-stability of PEI doped MoS$_2$ transistors' transfer curves were monitored by measuring the electronic performance as a function of time. Fig. 3(a) shows the exponential decay of drain current of 6, 36, 48, and 96 h after upon PEI molecules applied. To exclude the variability of each measurement and get accurate drain current, maximum ON-state drain current at $V_{bg} = 30$ V and $V_{ds} = 0.02$ V at certain time has been selected. Evolution of peak transconductance and ON/OFF ratio over time has also been shown in Fig. 3(b) to demonstrate the electronic behavior change over time. After the exposure to ambient air, the *n*-type behavior degraded due to the chemisorptions of oxygen and water molecules; after the first 6 h, the device experienced a stable degradation rate, but still exhibited good performance with high ON-current and extrinsic field-effect mobility. After additional time, the ON/OFF ratio had been recovered due to the degradation of PEI molecule layers. After four days exposure in ambient air, the FETs roughly return to the original undoped status. We also notice that the high performance can be once achieved by resoaking or spin coating the transistor with PEI solution.

## IV. CONCLUSION

In summary, we have demonstrated a strong *n*-type doping effect on MoS$_2$ FETs using PEI as a dopant source. It is capable of effectively changing the sheet and contact resistances of the MoS$_2$ flake as well as adjusting the threshold voltage. The large improvement of device performance in ON-current and extrinsic field-effect mobility is attributed to the reduction in sheet resistance and contact resistance due to the strong electron doping from PEI molecules. The method established in this letter could be widely applied to other TMD materials.